\documentclass[aps,twocolumn,superscriptaddress,showkeys,showpacs]{revtex4}
\usepackage{amssymb}
\usepackage{amsfonts}
\usepackage{amsmath}
\usepackage{amsthm}
\usepackage{epsfig}
\usepackage{color} 
\usepackage{subfigure}

\newcommand{\fr}[2]{\frac{\displaystyle #1}{\displaystyle #2}}
\newcommand{\df}[2]{\frac{\displaystyle d#1}{\displaystyle d#2}}

\begin{document}

\title{When Nonlocal Coupling Between Oscillators Becomes Stronger:\\ Patched Synchrony or Multi-Chimera States}

\author{Iryna Omelchenko}
\affiliation{Institut f{\"u}r Theoretische Physik, Technische Universit\"at Berlin, Hardenbergstra\ss{}e 36, 10623 Berlin, Germany}
\affiliation{Bernstein Center for Computational Neuroscience, Humboldt-Universit{\"a}t zu Berlin, Philippstra{\ss}e 13, 10115 Berlin, Germany}
\author{Oleh E. Omel'chenko}
\affiliation{Weierstrass Institute, Mohrenstra\ss{}e 39, 10117 Berlin, Germany}
\author{Philipp H{\"o}vel} 
\affiliation{Institut f{\"u}r Theoretische Physik, Technische Universit\"at Berlin, Hardenbergstra\ss{}e 36, 10623 Berlin, Germany}
\affiliation{Bernstein Center for Computational Neuroscience, Humboldt-Universit{\"a}t zu Berlin, Philippstra{\ss}e 13, 10115 Berlin, Germany}
\author{Eckehard Sch{\"o}ll}
\email[corresponding author: ]{schoell@physik.tu-berlin.de}
\affiliation{Institut f{\"u}r Theoretische Physik, Technische Universit\"at Berlin, Hardenbergstra\ss{}e 36, 10623 Berlin, Germany}

\date{\today}

\begin{abstract}
Systems of nonlocally coupled oscillators can exhibit complex spatio-temporal patterns,
called chimera states, which consist of coexisting domains of spatially coherent (synchronized) and 
incoherent dynamics. We report on a novel form of these states,
found in a widely used model of a limit-cycle oscillator
if one goes beyond the limit of weak coupling typical for phase oscillators.
Then patches of synchronized dynamics appear within the incoherent domain
giving rise to a multi-chimera state.
We find that, depending on the coupling strength and range,
different multi-chimeras arise in a transition from classical chimera states.
The additional spatial modulation
is due to strong coupling interaction
and thus cannot be observed in simple phase-oscillator models.
\end{abstract}

\pacs{05.45.Xt, 05.45.Ra, 89.75.-k}
\keywords{nonlinear systems, dynamical networks, coherence, spatial chaos}

\maketitle

During recent times the investigation of coupled systems has led to joint research efforts bridging between diverse fields such as nonlinear dynamics, network science, and statistical physics, with
a plethora of applications, e.g., in physics, biology, and technology. As the numerical resources have developed at fast pace, analysis and simulations of large networks with more and more sophisticated coupling schemes have come into reach giving rise to an abundance of new dynamical scenarios. Among these a very peculiar type of dynamics was first reported for the 
well-known model of phase oscillators. Such a network exhibits a hybrid nature combining both coherent and incoherent parts \cite{KUR02a,ABR04,LAI09,MOT10}, hence the name \textit{chimera states}. The most surprising aspect of this discovery was that these states exist
in a system of identical oscillators coupled in a symmetric ring topology with a symmetric interaction function. Recent works have shown that chimeras are not limited to phase oscillators, but can in fact be found in a large variety of different systems. These include time-discrete and time-continuous chaotic models \cite{OME11,OME12} and are not restricted to one spatial dimension. Also two-dimensional configurations allow for chimera states \cite{MAR10,OME12a}. Furthermore, similar scenarios exist for time-delayed coupling \cite{SET08}, and their dynamical properties and symmetries were subject to theoretical studies as well \cite{MAR10b,WOL11a,OME12}. It was only in the very recent past that chimeras were realized in experiments
on chemical oscillators \cite{TIN12} and electro-optical coupled-map lattices \cite{HAG12}.
The nonlocality of the coupling -- a crucial ingredient for chimera states -- also suggests an interesting connection to material science, see, for instance, magnetic Janus particles that undergo a synchronization-induced structural transition in a rotating magnetic field \cite{YAN12b,KLA12}. Nonlocality is of great importance not only for chimera states, but also for other dynamical phenomena such as turbulent intermittency \cite{BAT00}.
Hybrid states were also reported in the context of neuroscience under the notion of \textit{bump states} \cite{LAI01}. They were later confirmed for nonlocally coupled Hodgkin-Huxley models \cite{SAK06a} and 
may account for experimental observation of partial synchrony in neural activity during eye movement \cite{FUN91}.

In this letter, we present evidence that the habitat of chimeras indeed extends to neural models. This strongly suggests their universal appearance. Our findings also show that current knowledge of these hybrid states is far from being complete: Next to the classical chimera state, which exhibits one coherent, phase-locked and one incoherent region, we find a new class of dynamics that possesses multiple domains of incoherence.

We consider a ring of $N$ nonlocally coupled FitzHugh-Nagumo (FHN) oscillators, whose relevance is not limited to neuroscience, but also includes chemical \cite{SHI04} and optoelectronic \cite{ROS11a} oscillators and nonlinear electronic circuits \cite{HEI10}:
\begin{subequations}\label{System:FHN}
\begin{align}
\varepsilon \frac{d u_k}{dt} & = u_k - \fr{u_k^3}{3} - v_k \nonumber\\
                     & + \frac{\sigma}{2R} \sum\limits_{j=k-R}^{k+R}
\left[ b_{\mathrm{uu}}( u_j - u_k ) + b_{\mathrm{uv}}( v_j - v_k ) \right],\\
\frac{d v_k}{dt} &= u_k + a_k \nonumber\\
                     & + \frac{\sigma}{2R} \sum\limits_{j=k-R}^{k+R}
\left[ b_{\mathrm{vu}}( u_j - u_k ) + b_{\mathrm{vv}}( v_j - v_k ) \right].
\end{align}
\end{subequations}
where $u_k$ and $v_k$ are the activator and inhibitor variables, respectively \cite{FIT61,NAG62},
and $\varepsilon>0$ is a small parameter characterizing a timescale separation, which we fix at $\varepsilon = 0.05$ throughout the paper.
Depending upon the threshold parameter $a_k$, each individual FHN unit exhibits
either oscillatory ($|a_k|<1$) or excitable ($|a_k|>1$) behavior.
All indices are modulo $N$.
In this study we assume that the elements are in the oscillatory regime and identical, i.e., $a_k \equiv a\in(-1,1)$.
 The form of the coupling in Eqs.~(\ref{System:FHN}) is inspired from neuroscience \cite{KOZ98,VAN04a,HEN11}. Neuronal networks are often structured in topologies where strong interconnections between different neurons
are found within a range~$R$, but much fewer connections exist at longer distances (see Ref.~\cite{HEN11} and references therein). We approximate this feature by constant coupling with a strength~$\sigma>0$
within the~$R$ nearest neighbors from both sides, and vanishing coupling at longer distances. We stress that chimera states have previously been found for various
types of nonlocal coupling \cite{ABR04,KUR02a,OME11,HAG12}, including exponential coupling functions which can be derived from adiabatic elimination of an additional fast, diffusing variable \cite{SHI04}. But the crucial common feature of all these coupling functions is the non-locality \cite{footnote1}. 

As we will demonstrate,  another important feature of Eqs.~(\ref{System:FHN})
is that it contains not only direct $u$-$u$ and $v$-$v$ couplings,
but also cross-couplings between activator ($u$) and inhibitor ($v$) variables.
For the sake of simplicity, we model this feature
by a rotational coupling matrix
\begin{equation}
B = \left(
\begin{array}{ccc}
b_{\mathrm{uu}} & & b_{\mathrm{uv}} \\
b_{\mathrm{vu}} & & b_{\mathrm{vv}}
\end{array}
\right) =
\left(
\begin{array}{ccc}
\cos \phi  & & \sin \phi \\
-\sin \phi  & & \cos \phi
\end{array}
\right)
\label{Matrx:B}
\end{equation}
depending on a single parameter $\phi\in[-\pi,\pi)$. Thus we can vary four control parameters
of different nature: $a$ determining the local dynamics, $\sigma$, $R$, and $\phi$ specifying the coupling.

\begin{figure}[t!]
\includegraphics[width=\linewidth]{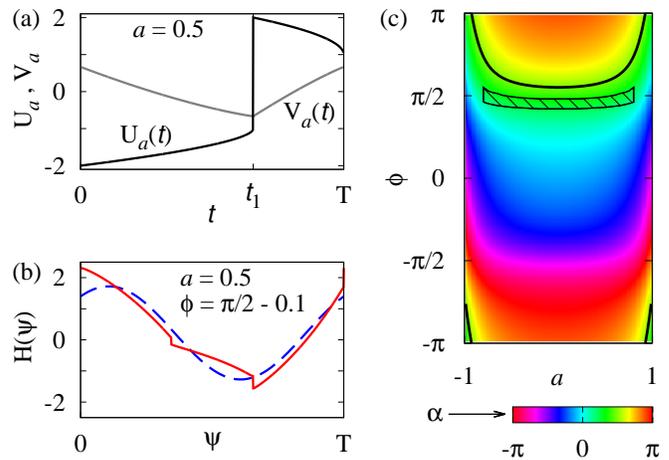}
\caption{(Color online) (a) Limit cycle $u=U_a(t),v=V_a(t)$ of single decoupled FHN oscillator for $\varepsilon\to 0$.
(b) Phase interaction curve~$H(\psi)$ (red solid) obtained from Eq.~(\ref{Def:H})
and its approximation (blue dashed) by the partial Fourier sum~(\ref{Approximation:Alpha}).
(c) Phase lag parameter~$\alpha$ as a function of control parameters~$a$ and~$\phi$.
The hatched region indicates the parameter range for which we find chimera states
in simulations of Eqs.~(\ref{System:FHN}) with $N=1000$, $R = 350$ and $\sigma = 0.05$.
The thick black curve for $\alpha = \pi / 2$ is a guide for the eye.}
\label{fig:H}
\end{figure}

In the following we address the question of appropriate choice of the coupling phase $\phi$ by a phase-reduction technique in the limit $\varepsilon\to 0$. In the absence of coupling the local dynamics is then periodically oscillating and follows a limit 
cycle~$\left(u=U_a(t),v=V_a(t)\right)$,
see Fig.~\ref{fig:H}(a), with the approximate period \cite{BRA09}
\begin{equation}
T = 3 + (1-a^2) \ln\left( \fr{1-a^2}{4-a^2} \right)
\end{equation}
and two discontinuities at times
$$
t_1 = \fr{3}{2} + a + (1-a^2) \ln\left( \fr{1-a}{2-a} \right)\quad\mbox{and}\quad t_2 = T.
$$
Here, $U_a(t)$ is composed of two distinct solutions of the differential equation
on the slow activator nullcline \cite{BRA09}
\begin{equation}
\df{u}{t} = \frac{u + a}{1 - u^2}.
\end{equation}
The first is defined for $t\in[0,t_1)$ and satisfies initial condition $U_a(0)=-2$,
while the second is defined for $t\in[t_1,T)$ and starts from the point $U_a(t_1)=2$.
The corresponding slow $v$-variable is given by $V_a(t) = U_a(t) - U_a^3(t)/3$.
In the weak coupling limit $\sigma\ll\varepsilon\ll 1$,
we now perform a phase reduction of Eqs.~(\ref{System:FHN}) \cite{IZH00b}.
Then the dynamics of each FHN oscillator is conveniently described
by a single scalar phase variables $\theta_k\in \mathbb{R}\, \mathrm{mod}\, T$,
which reflects the position of this oscillator along its unperturbed limit cycle.
The interaction of these phases, to the leading order, is given by a reduced system
\begin{equation}
\frac{d\theta_k}{dt} = - \frac{1}{2R} \sum\limits_{j=k-R}^{k+R} \left[ H( \theta_k - \theta_j ) - H(0) \right],
\label{PhaseModel}
\end{equation}
where the function
\begin{align}
H(\psi) & = \frac{1}{T} \int_0^T \frac{ p(t - \psi) - \left( 1 - U_a^2(t) \right) q(t - \psi) }{ \left( 1 - U_a^2(t) \right) ( U_a(t) + a ) } dt\nonumber\\
            & -\frac{1}{T} \sum\limits_{m=1}^2 c_m p(t_m - \psi)\label{Def:H}
\end{align}
is $T$-periodic and we use the abbreviations
\begin{subequations}
\begin{align}
p(t) & = U_a(t) \cos\phi + V_a(t) \sin\phi,\\
q(t) & = - U_a(t) \sin\phi + V_a(t) \cos\phi,\\
c_1 &= -\frac{3}{(a-1)(a+2)},\quad c_2=\frac{3}{(a+1)(a-2)}
\end{align}
\end{subequations}
for notational convenience.
The function $H(\psi)$ typically has several discontinuities (see solid (red) curve in Fig.~\ref{fig:H}(b)),
because the FHN model becomes a relaxation oscillator for $\varepsilon\to 0$.

\begin{figure}[t!]
\includegraphics[height=\linewidth, angle=270]{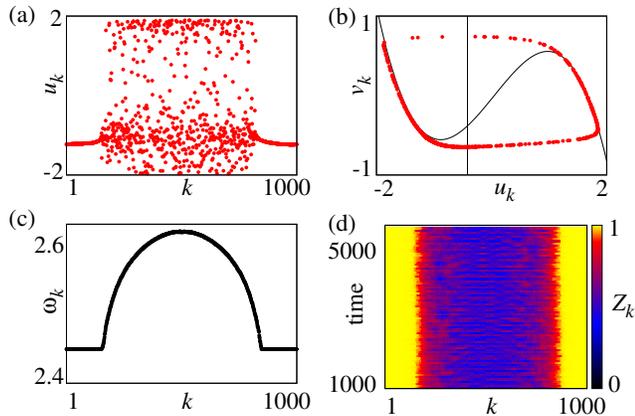}
\caption{(Color online) (a) Snapshot of the variables $u_k$ for $t=5000$, (b)~snapshot in the $(u_k,v_k)$-plane for $t=5000$ (black lines denote the nullclines of the FHN system), (c)~mean phase velocities $\omega_k$, (d)~local order parameter $Z_k$. Parameters: $N=1000$, $r=0.35$, $\sigma=0.1$, $a=0.5$, $\phi= \pi/2-0.1$.
}
\label{fig2}
\end{figure}

In order to find suitable parameter values for the generation of chimera states,
we employ results of Ref.~\cite{OME10a}.
There it was shown that chimera states can be generically found
in systems of the form of Eq.~(\ref{PhaseModel}) with $H(\psi) = \sin(\psi + \alpha)$
if the phase lag parameter $\alpha$ is close to but less than $\pi/2$.
Phase interaction curves $H(\psi)$ corresponding to a FHN oscillator
can be qualitatively approximated by their Fourier series truncated at the first order
\begin{eqnarray}
H(\psi) &\approx& \fr{h_0}{2} + h_1^{\mathrm{c}} \cos\left( \fr{2\pi}{T} \psi \right) + h_1^{\mathrm{s}} \sin\left( \fr{2\pi}{T} \psi \right)\nonumber \\
&=& \fr{h_0}{2} + \sqrt{ \left( h_1^{\mathrm{c}} \right)^2 + \left( h_1^{\mathrm{s}} \right)^2 } \sin\left( \fr{2\pi}{T} \psi + \alpha \right),
\label{Approximation:Alpha}
\end{eqnarray}
where the Fourier coefficients $h_0$, $h_1^c$, and $h_1^s$ are given by
\begin{subequations}
\begin{align}
h_0              & = \frac{2}{T} \int_0^T H(\psi) d\psi,\\
h_1^{\mathrm{c}} & = \frac{2}{T} \int_0^T H(\psi) \cos\left( \fr{2\pi}{T} \psi \right) d\psi,\\
h_1^{\mathrm{s}} & = \frac{2}{T} \int_0^T H(\psi) \sin\left( \fr{2\pi}{T} \psi \right) d\psi.
\end{align}
\end{subequations}
This yields the following approximate equations for the phase-lag parameter $\alpha$
\begin{align}
\cos\alpha = \fr{h_1^{\mathrm{s}}}{\sqrt{ \left( h_1^{\mathrm{c}} \right)^2 + \left( h_1^{\mathrm{s}} \right)^2 }},\quad
\sin\alpha = \fr{h_1^{\mathrm{c}}}{\sqrt{ \left( h_1^{\mathrm{c}} \right)^2 + \left( h_1^{\mathrm{s}} \right)^2 }},
\end{align}
that can be used to pinpoint a region in the parameter plane $(a,\phi)$,
which favors the appearance of chimera states.
Roughly speaking, such states are expected for a pronounced off-diagonal coupling ($\phi \approx \pi/2$),
but not for a diagonal one ($\phi \approx 0$ or $\phi \approx \pi$).
This prediction is confirmed by numerical simulations, see the hatched area in Fig.~\ref{fig:H}(c).
Indeed, if we choose~$a=0.5$ and~$\phi = \pi/2 - 0.1$,
then for~$\sigma$ small enough we obtain a chimera-like solution shown in Fig.~\ref{fig2},
where we used initial conditions randomly distributed on the circle $u^2 + v^2 = 4$.

Figure~\ref{fig2}(a) shows a snapshot of variables $u_k$ at time $t=5000$.
One can clearly distinguish coherent and incoherent parts, a characteristic signature of chimera states. 
Elements that belong to the incoherent part are scattered along the limit cycle, see Fig.~\ref{fig2}(b).
The subsystems of this region perform a nonuniform rotational motion, but neighboring oscillators are not phase-locked. To illustrate this, Fig.~\ref{fig2}(c) shows mean phase velocities for each oscillator calculated as $\omega_k = 2 \pi M_k/\Delta T$, $k=1,...,N$,
where $M_k$ is the number of complete rotations around the origin performed by the $k$-th unit during the time interval $\Delta T$. The values of $\omega_k$ lie on a continuous curve and the interval of constant $\omega_k$ corresponds to the coherent region, where neighboring elements are phase-locked. This phase velocity profile is a clear indication of chimera states and similar to the case of coupled phase oscillators.

The spatial coherence and incoherence of the chimera state can be characterized by a real-valued local order parameter \cite{WOL11a,OME11}
\begin{align}
Z_k = \left| \frac{1}{2\delta} \sum_{|j-k|\leq\delta}e^{i\Theta_j}\right|, ~~~ k=1,\dots,N,
\end{align}
where  $\Theta_j= \arctan(v_j/u_j)$ denotes the geometric phase of $j$-th FHN unit. We use a spatial average with a window size of $\delta=25$ elements. A local order parameter $Z_k=1$ indicates that the $k$-th unit belongs to the coherent part of the chimera state, and $Z_k$ is less than $1$ for incoherent parts. 
Figure~\ref{fig2}(d) depicts the local order parameter in the time interval $t\in[1000,5000]$, where bright (yellow) color denotes the coherent regions.

\begin{figure}[t!]
\includegraphics[height=\linewidth, angle=270]{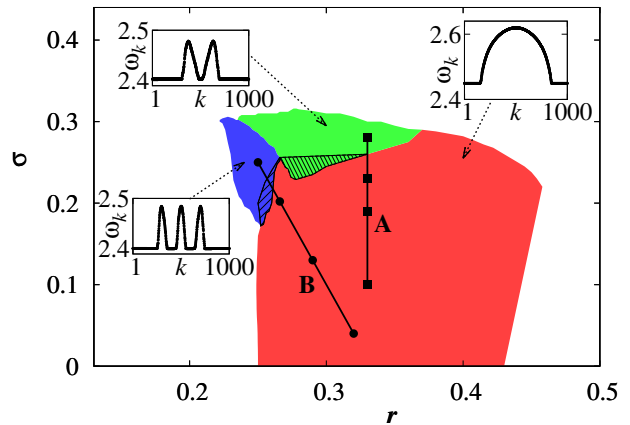}
\caption{(Color online) Stability regimes for chimera states with one (gray, red), two (dark-gray, green), and three (blue, black) incoherent domains in the plane of coupling radius $r$ and coupling strength $\sigma$. Other parameters as in Fig~\ref{fig2}. Hatched regions denote multistable regimes. Insets show typical profiles of the mean phase velocities. Black squares (A) and circles
(B) denote parameter values for the transition scenarios shown in Fig.~\ref{fig4}.}
\label{fig3}
\end{figure}

\begin{figure*}[Ht!]
\includegraphics[width=\linewidth]{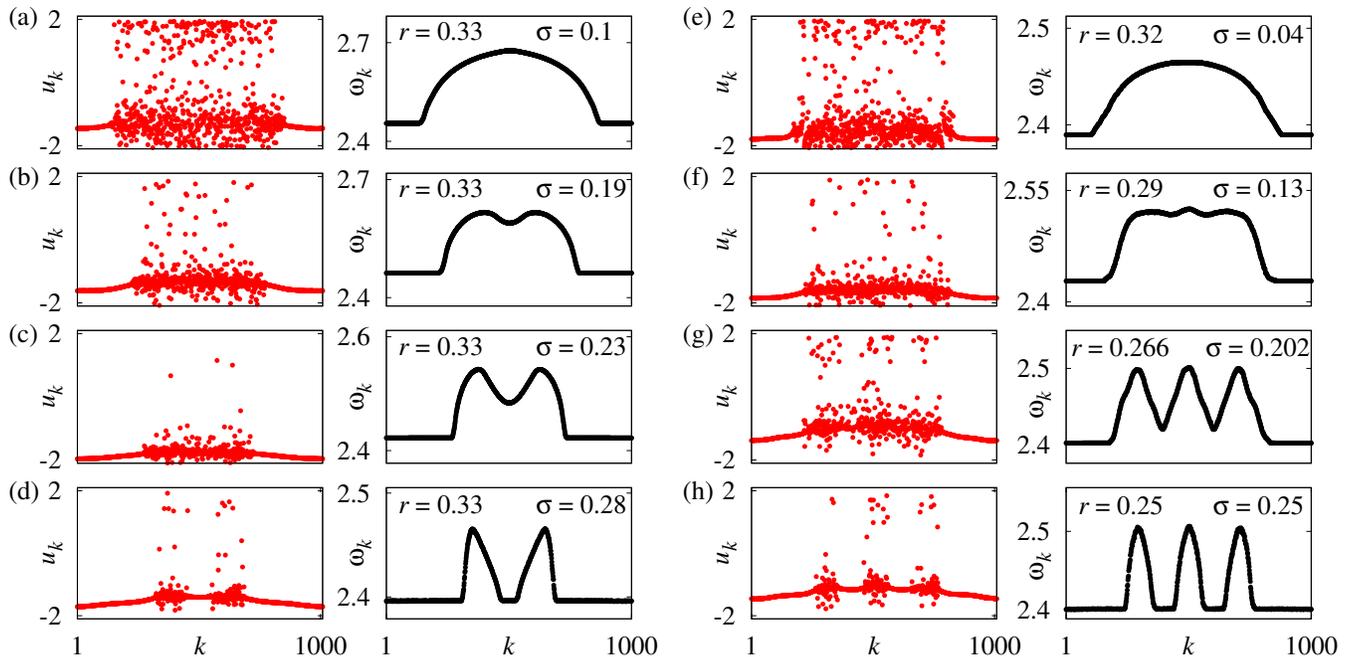}
\caption{(Color online) Transition from a classical chimera state with one incoherent domain to multi-chimera states with two~(a)-(d), and three~(e)-(h) incoherent domains.
In each panel the left column shows snapshot of variables $u_k$, and the right column shows the corresponding mean phase velocities. Coupling radius and strength follow the black squares (A) in Fig.~\ref{fig3} for panels~(a)-(d) and black circles (B, $\sigma=-3r+1$) in Fig.~\ref{fig3} for panels~(e)-(h).
Other parameters as in Fig.~\ref{fig2}.}
\label{fig4}
\end{figure*}

For further investigation we fix the values of parameters~$a=0.5$ and~$\phi=\pi/2-0.1$ and vary 
radius~$r$ and strength~$\sigma$ of the coupling.
In previous works, where nonlocally coupled phase oscillators,
discrete maps, R{\"o}ssler and Lorenz systems were considered, 
chimera states were reported for intermediate coupling radii and small coupling strengths \cite{OME11,OME12}. Figure~\ref{fig3} displays the stability diagram for chimera states
in nonlocally coupled FHN systems in a similar parameter range. The gray (red) region corresponds to the classical chimera state with one incoherent domain. Surprisingly, for increasing coupling strength, we observe qualitatively new types of chimera states, which have two or even three incoherent domains indicated by the dark-gray (green) and black (blue) regions, respectively. We call these states \textit{multi-chimera states}. Their additional spatial modulation cannot be explained in terms of phase interaction only \cite{SET08}.
Near the borders of the different regimes, multistability is found and indicated by hatching. In this region of coexistence of one and two (or one and three) incoherent domains the particular realization depends on the choice of initial conditions. 

The transition from the classical to the multi-chimera state is illustrated in Fig.~\ref{fig4}. Following the lines with black squares (A) and black circles (B) in Fig.~\ref{fig3}, Figs.~\ref{fig4}(a)-(d) and \ref{fig4}(e)-(h) demonstrate how multi-chimeras with two and three incoherent domains develop from the classical chimera with one incoherent part. The velocity profile shows a dip, which becomes more pronounced as $\sigma$ increases and eventually reaches down to the level of the coherent part. For smaller coupling radii, Eq.~(\ref{System:FHN}) exhibits increasing multistability. There we also find chimera states with more than three incoherent parts, but these states have relatively small stability domains and are visible for large system sizes~$N$ only.

In conclusion, we have reported the existence of chimera states for relaxation oscillators,
which are of slow-fast type and exhibit a timescale separation between activator and inhibitor.
Applying a phase-reduction technique, we have identified nonlocal off-diagonal coupling
to be a crucial ingredient for the occurrence of these hybrid states that exhibit coexistent coherent and incoherent domains. Our findings corroborate the universal existence of chimera states,
which were previously reported for phase oscillators and time-discrete or time-continuous chaotic systems. Furthermore, we have found a new type of multi-chimera states that consist of multiple domains of incoherence. They appear as a result of strong coupling interaction and thus cannot be found in simple phase models.
Such multi-chimera states are of generic nature and can also be found
for other non-local coupling functions, see Supplemental Material.
The fact that multi-chimeras with increasing numbers of incoherent domains appear successively with decreasing range of coupling (cf. Fig.~\ref{fig3}) is reminiscent of the appearance of successive coherence tongues of increasing wavenumbers 
in rings of time-discrete or time-continuous chaotic systems, where it has been related to the scaling behavior of the spatial profiles with rescaled space variable and coupling range \cite{OME12,HAG12}. This will be a promising direction of future research.

IO and PH acknowledge support by the Federal Ministry of Education and Research (BMBF), Germany (grant no. 01GQ1001B).
This work was also supported by DFG in the framework of SFB 910.


\clearpage
\onecolumngrid

\section*{Supplemental material}

Consider a ring of FitzHugh-Nagumo elements with an exponential coupling function:
\begin{eqnarray}
\varepsilon \frac{d u_k}{dt} & = & u_k - \frac{u_k^3}{3} - v_k + \frac{\sigma}{N} \sum\limits_{j=1}^N G_{kj}
\left[ b_{\mathrm{uu}}( u_j - u_k ) + b_{\mathrm{uv}}( v_j - v_k ) \right],\nonumber\\
\frac{d v_k}{dt} & = & u_k + a + \frac{\sigma}{N} \sum\limits_{j=1}^N G_{kj}
\left[ b_{\mathrm{vu}}( u_j - u_k ) + b_{\mathrm{vv}}( v_j - v_k ) \right],
\end{eqnarray}
where we used the abbreviations
\begin{equation}
B = \left(
\begin{array}{ccc}
b_{\mathrm{uu}} & & b_{\mathrm{uv}} \\
b_{\mathrm{vu}} & & b_{\mathrm{vv}}
\end{array}
\right) =
\left(
\begin{array}{ccc}
\cos \phi  & & \sin \phi \\
-\sin \phi  & & \cos \phi
\end{array}
\right)
\end{equation}
and
\begin{equation}
G_{kj} = C_\kappa \exp\left( - \frac{ \kappa | k - j | } { N } \right),\quad\mbox{with}\quad C_\kappa = \frac{\kappa}{2 (1 - e^{-\kappa/2})}.
\end{equation}
Here, discrete distance~$|k - j|$ is evaluated with respect to ring topology,
and~$C_\kappa$ is a normalization constant.

\begin{figure}[h!]
\centering
\includegraphics[width=0.5\linewidth, angle=270]{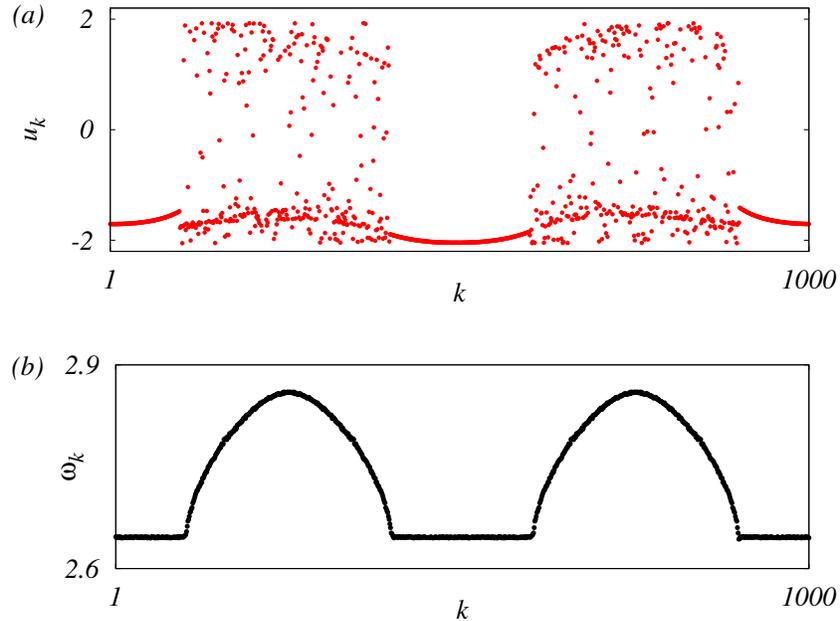}
\caption{Multi-chimera state with exponential coupling function:
(a) Snapshot of $u_k$ for $t=1000$. (b) Mean phase velocities $\omega_k$.
Parameters: $a=0.5$, $\varepsilon=0.05$, $N=1000$, $\kappa = 4$, $\sigma = 0.25$, and $\phi = \pi / 2 -0.1$.}
\label{suppl_fig1}
\end{figure}

Figure~\ref{suppl_fig1} shows a multi-chimera state with two regions of spatial coherence and incoherence for this exponential coupling.

\end{document}